\documentclass[a4paper,11pt]{article}

\usepackage{pos}

\usepackage{lineno}

\title{Prospects for sub-GeV astrophysical neutrino detection with IceCube}

\ShortTitle{Sub-GeV IceCube}

\author{The IceCube Collaboration \\{\normalsize \normalfont(a complete list of authors can be found at the end of the proceedings)}\\}

\emailAdd{per.myhr@uclouvain.be}
\emailAdd{gwenhael.dewasseige@uclouvain.be}

\abstract{
The IceCube Neutrino Observatory is currently the largest and most sensitive detector for astrophysical neutrinos and has pioneered the field of high-energy neutrino astronomy. Despite being designed with the primary goal of identifying astrophysical TeV neutrinos and their corresponding sources, recent studies, utilising the DeepCore subdetector, have shown IceCube's proficiency in being sensitive to astrophysical neutrinos at GeV energies. Currently, there is a gap in sensitivity between the supernova detection system at MeV energies and the lowest-energy triggering events around 1 GeV. In this contribution, we present the ongoing efforts to cover this gap and increase the sensitivity of IceCube to sub-GeV astrophysical neutrinos. Despite high background rates, we show how the complimentary use of manifold and supervised machine learning can make IceCube sensitive to neutrinos from transient sources down to energies of 100 MeV.

\vspace{4mm}

{\bfseries Corresponding authors:}
Per Arne Sevle Myhr$^{1*}$, 
Gwenhaël de Wasseige$^{1}$\\

{$^{1}$ \itshape Centre for Cosmology, Particle Physics and Phenomenology,\\
Université catholique de Louvain}\\[4mm]
$^*$ Presenter
}

\ConferenceLogo{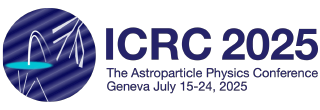}

\FullConference{39th International Cosmic Ray Conference (ICRC2025)\\
 15–24 July 2025\\
Geneva, Switzerland\\}

\begin{document}

\maketitle

\section{Introduction: Astrophysical sub-GeV neutrinos}\label{sec1}
Neutrino astronomy is a relatively young field of research, especially compared to its electromagnetic counterpart. Because of their minuscule interaction cross-section compared to other elementary particles, neutrinos can pass through gaseous regions and carry information from regions otherwise opaque to electromagnetic radiation. Furthermore, they are unaffected by electric and magnetic fields, avoiding both attenuation and deflections as they propagate over cosmological distances. These attributes make neutrinos great probes for the physical processes inside energetic environments, which cannot be directly accessed through electromagnetic observations. As the observed direction of astrophysical neutrinos coincides with the direction of their source and as they travel uninterrupted over vast distances, they play an important part in the growing field of multi-messenger astronomy. Neutrinos offer a unique probe to the particle physics at play in astrophysical environments, and the identification of astrophysical neutrino sources is a strong indication of hadronic processes. Over the last decade, detection of high-energy astrophysical neutrinos has both confirmed their existence~\cite{IceCube:2014stg} and provided an indirect probe to particle physics at energies not reachable at Earth.

Astrophysical neutrinos are produced over a wide range of energies. The first positive detection of extraterrestrial neutrinos was from the p--p fusion processes in the core of the Sun \cite{Davis:1994jw}. Similarly energetic neutrinos are produced in supernovae, where $>99\%$ of the released energy is contained in $\sim$ MeV neutrinos \cite{Mirizzi:2015eza}. Neutrinos from a local core-collapse supernova were detected by the Kamiokande II and other experiments in 1987 \cite{Kamiokande-II:1987idp, Bionta:1987qt, Alekseev:1988gp}. At higher energies, neutrinos with TeV--PeV energies have been detected by the IceCube Neutrino Observatory \cite{IceCube:2016zyt} and a few sources have been identified. TXS 0506+056 was the first neutrino source identified by coincident electromagnetic observations \cite{IceCube:2018dnn}, followed by the more recent discovery of NGC 1068 \cite{IceCube:2022der} and the Galactic Plane \cite{IceCube:2023ame}. Neutrinos from these known sources are expected to be created by the interaction of cosmic ray protons with surrounding photon fields. Neutrinos between $\sim$10 MeV and TeV are instead thought to be produced in proton--proton and proton--neutron collisions \cite{Kelner:2006tc}. IceCube has already performed multiple searches for GeV-scale neutrinos from transient astrophysical objects where proton acceleration is expected, but so far only upper limits have been established \cite{Abbasi:2022whi}.

The current global landscape of astrophysical neutrino detection consists of a handful of detectors. In addition to IceCube, data from other gigaton-scale detectors currently under construction \cite{KM3Net:2016zxf, Baikal-GVD:2018isr}, and other types of detectors, such as the Super-Kamiokande experiment, provide complementary sensitivity to astrophysical neutrinos down to MeV energies. In IceCube, there exists a more densely instrumented sub-array of the detector called DeepCore, designed to improve the sensitivity to GeV-scale neutrinos \cite{IceCube:2011ucd}. Due to the higher density and efficiency of the DeepCore photomultiplier tubes, DeepCore has been used to detect atmospheric neutrinos and study the properties of neutrino oscillations \cite{IceCubeCollaboration:2023wtb}. Besides its advancements in oscillation physics, DeepCore improves IceCube's sensitivity to GeV-scale astrophysical neutrinos from transient objects by comparing the steady background rates, dominated by atmospheric and detector noise, with the expected increase in the astrophysical neutrino flux over a short time window. In the following sections, we propose to use sub-threshold IceCube data to search for astrophysical neutrinos with $\mathcal{O}$(GeV) energies.

\section{Subthreshold data in IceCube}\label{sec2}
The IceCube detector consists of a tridimensional matrix of 5160 digital optical modules (DOMs) arranged 60 at a time over 86 strings buried up to 2.5 km in the South Pole ice. All 60 DOMs on a single string are connected to a DOMHub at the surface, which handles both power transmission to the DOMs and all data transfer to and from the DOMs. One step abstracted from the individual DOMHubs is the main surface data acquisition (pDAQ) system. After synchronising data from all 86 strings, pDAQ implements a trigger system to which all raw data is sent. The most general triggers look for a certain number of minimal multiplicity hits (i.e., hits on neighbouring DOMs within the light-travel time between them) within some predefined geometry and time window. Multiple different triggers are run in parallel and are combined further downstream by a global trigger algorithm that continuously reads information from all triggers. This global trigger then sends the combined trigger requests to an event builder that extracts the information about which DOMs have been hit and at what time and requests the entire IceCube detector to be read out in a time-window centred on the time of the global trigger. Finally, the full detector data is sent to a processing and filtering (PnF) system running on computers hosted in the lab at the surface. Due to the limited bandwidth available to transfer data from the South Pole to the north, further filters are applied at this stage, and only events passing these filters are transferred via satellite. These steps make up a rough outline of how data is acquired at the South Pole.
\begin{figure}[hbtp]
    \centering
    \includegraphics[width=0.7\linewidth]{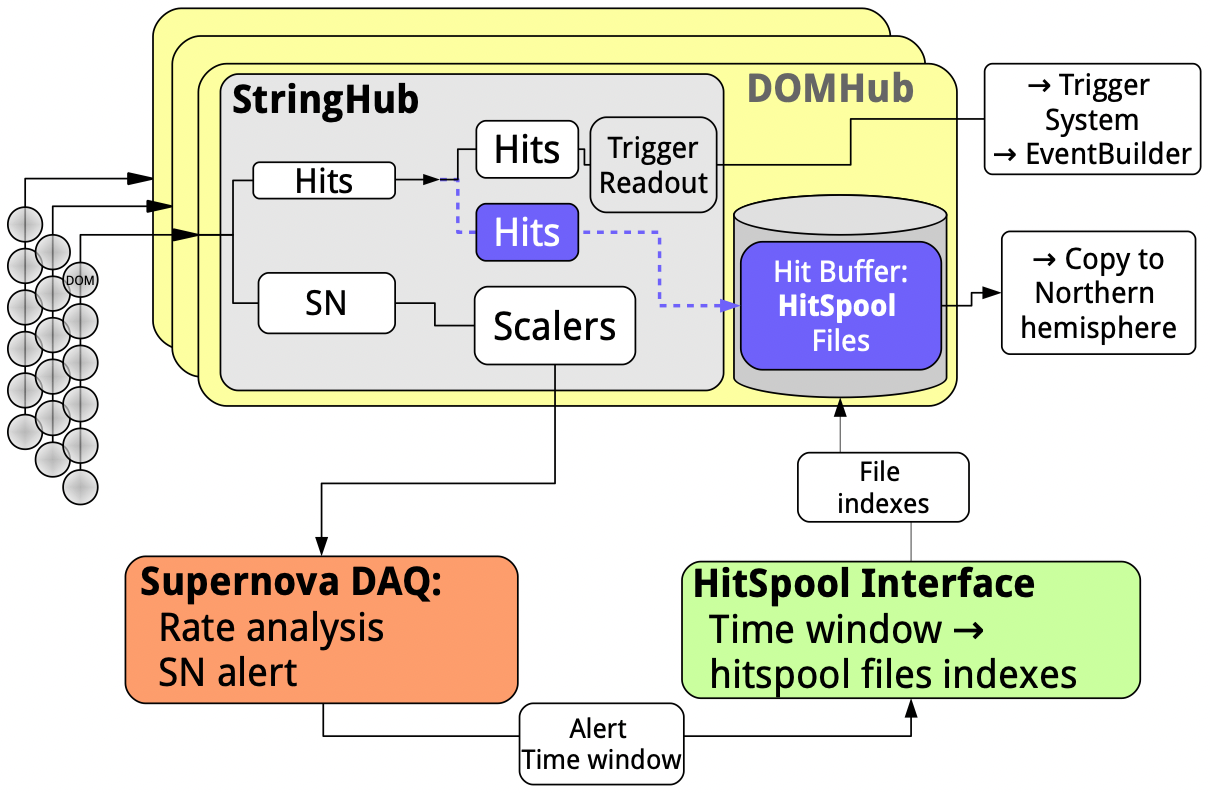}
    \caption{Flowchart of how data is processed from the DOMs on the strings until an event is made and sent by satellite to the north. HitSpool data is copied into a buffer in the DOMHubs prior to any trigger readout and is temporarily saved for 12.5 days (blue boxes). Image from \cite{HeeremanvonZuydtwyck:2015mbs}.}
    \label{fig:daq_data_flow}
\end{figure}
\subsection{HitSpooling}
HitSpool is a buffer of raw data sampled from an early stage of the processing chain summarised above, designed with the primary goal of improving data acquisition relevant for supernova detection in IceCube. A copy of all data from each string is made before it reaches the DOMHub, where all hits are spooled into a temporary data buffer. This buffer can store unprocessed data for almost 14 days and allows for a posteriori requests to save the full detector output through a live system. Data in this format is saved for all significant public alerts from the LIGO-Virgo-KAGRA gravitational wave interferometers. For a more thorough description of the HitSpooling system, see ref. \cite{HeeremanvonZuydtwyck:2015mbs}.

\subsection{Sub-threshold data processing pipeline}
Until recently, HitSpool data was mainly used for different supernova-related analyses with IceCube. However, we have implemented a new data processing pipeline that allows us to work with HitSpool data within the existing software framework used within IceCube. This pipeline allows for further reduction of data to extract sub-threshold events, events that would not result in a trigger in the regular processing scheme. It is preferred over implementing a new filter because sub-GeV neutrinos may not even initiate any of the triggers. Using this framework, we can start searching for a potential astrophysical neutrino signal in otherwise discarded data.
 
Since HitSpool data contains every single hit in the entire detector, we perform additional cuts to reduce this noise-dominant data down to manageable levels. These cuts are summarised as follows:
\begin{itemize}
    \item \textit{Discard data that does not contain any DOM-to-DOM multiplicity hits.} The HitSpool data is divided into discrete packs of raw data each lasting for $\sim$ 100 $\mu s$. If no neighbouring DOMs have seen hits within a pre-defined time, related to the light-travel time between them, then this data package is discarded. This step removes $\sim90\%$ of the data.
    \item \textit{Filter out any data packages that would cause a trigger in the main data acquisition.} Since we are interested in events that are sub-threshold, events passing any of the standard triggering algorithms implemented in pDAQ are not interesting to us. This data is available without the use of HitSpool and is, by definition, not sub-threshold. Only a small fraction of data is removed in this step.
    \item \textit{Apply sub-threshold muon-rejection algorithms.} We use previously developed HitSpool data algorithms that categorise low-energy atmospheric muons -- muons created by cosmic-ray interactions in the atmosphere -- that are low enough in energy to not trigger in IceCube. Despite the background at GeV energies being completely dominated by noise from the detector and surrounding environment, atmospheric muons can still contaminate our sample. This step filters about $50\%$ of the remaining data.
\end{itemize}
 
By implementing the three simple steps described above, we are left with a large set of sub-threshold data ready for higher-level analysis.

\section{Identifying sub-threshold events}\label{sec3}
To search for correlated hits in the reduced HitSpool data, a new algorithm has been developed that looks for a collection of hits correlated in space and time. As this method searches for a small cluster or ‘burst' of hits caused by the interaction of a single low-energy neutrino, it is refereed to as a `burst-search' algorithm in the following.

\subsection{Burst parameters}
\begin{figure}[hbtp]
    \centering
    \includegraphics[width=0.49\linewidth]{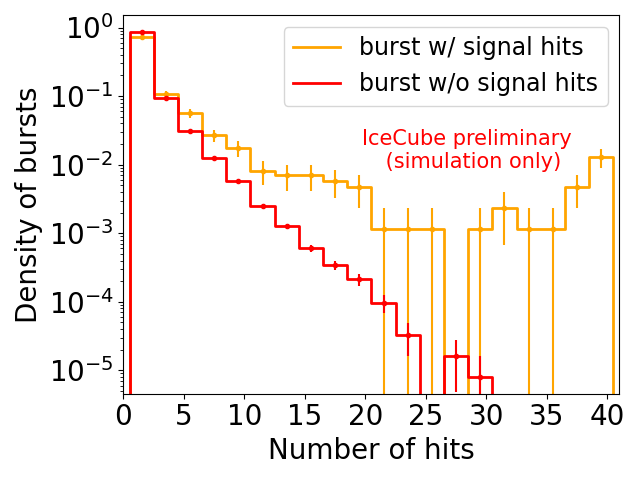}
    \includegraphics[width=0.49\linewidth]{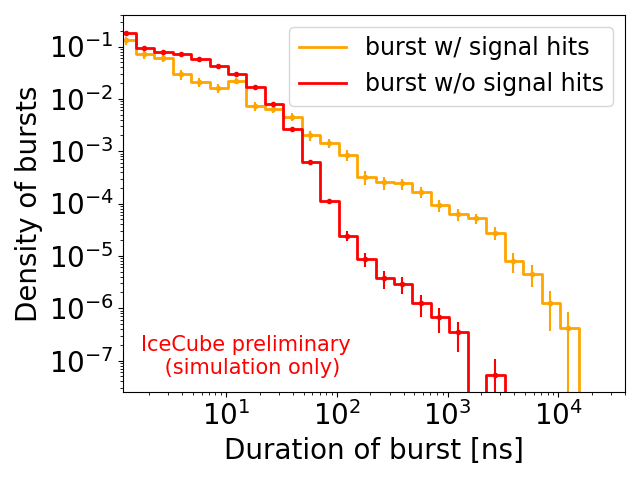}
        \caption{\textit{Left:} Density distribution of the number of hits in a burst, referred to as the size of the bursts, from simulations. The orange distribution shows bursts containing simulated muon neutrinos, whereas the red distribution shows bursts containing only noise. The statistical fluctuations of each bin are shown with the error bars. \textit{Right:} Density distribution of the duration of the bursts from simulation. Colours are the same as in the left plot.}
    \label{fig:burst_parameter_distribution}
\end{figure}

The main goal of the burst-search algorithm is to pick up on small collections of hits inside the detector characteristic of the signal we expect from $\mathcal{O}$(GeV) neutrinos. It relies on three initial parameters estimated from simulations:
\begin{itemize}
    \item Time difference between hits: This parameter is set to $\Delta t\leq 3\;\mu s$.
    \item Distance between pair-wise hits: Pairs of hits must be $D\leq 200\;m$.
    \item Minimal size of burst: The minimal number of hits that make up a burst. This parameter is set to $S\geq 2$.
\end{itemize}
The algorithm works by ordering all hits inside the detector in time, then running a sliding time-window over these hits, combining hits within the distance and time mentioned above into bursts. With these three initial parameters defined, we run the burst-search algorithm on simulated sub-threshold events -- simulated neutrinos with energies between 100 MeV and 1 GeV that do not cause a trigger -- and HitSpool data. Some of the resulting parameter distributions are shown in Figure \ref{fig:burst_parameter_distribution}.

In Figure \ref{fig:burst_parameter_distribution}, the signal and background distributions show clear differences. In the left panel, the majority of bursts picked up by the algorithm are smaller bursts of 2 or 3 hits. However, as we move towards bursts of larger size, the distributions diverge. We see significantly more bursts with large number of hits when considering those containing simulated muon neutrinos. Additionally, we see from the right panel of Figure \ref{fig:burst_parameter_distribution} that bursts with muon neutrinos show a tendency to last for a longer duration than those without. The difference in the tail of the two distributions shows the potential of this algorithmic burst-search approach to differentiate between GeV neutrinos and background on a statistical level.
 
From the output of the burst-search algorithm, we compute a set of new values for each burst: \textit{frequency} -- average number of hits in the burst per $\mu s$, \textit{charge} -- the total number of photoelectrons deposited in the burst (that may be used as a proxy for the deposited energy), \textit{hyperplane} -- the projected area of the hits in the burst along the z-axis, \textit{50\% quantile} -- time for half of the hits to occur within a burst, \textit{strings} -- number of strings in the burst, and \textit{CoG} -- the charge-weighted centre of gravity along each x, y, and z coordinate. Together with the initial parameters of duration and size, these values make up a higher-dimensional representation of the burst for further analysis.

\subsection{Manifold learning}
To see if any of the additional parameters introduced above are successful in distinguishing between simulated neutrinos and background, we perform a principal component analysis (PCA) on a large collection of bursts. The resulting principal components show no clear clustering of data points, and linear dimensionality reduction is insufficient for distinguishing neutrinos from different backgrounds.
 
\begin{figure}[hbtp]
    \centering
    \includegraphics[width=0.45\linewidth]{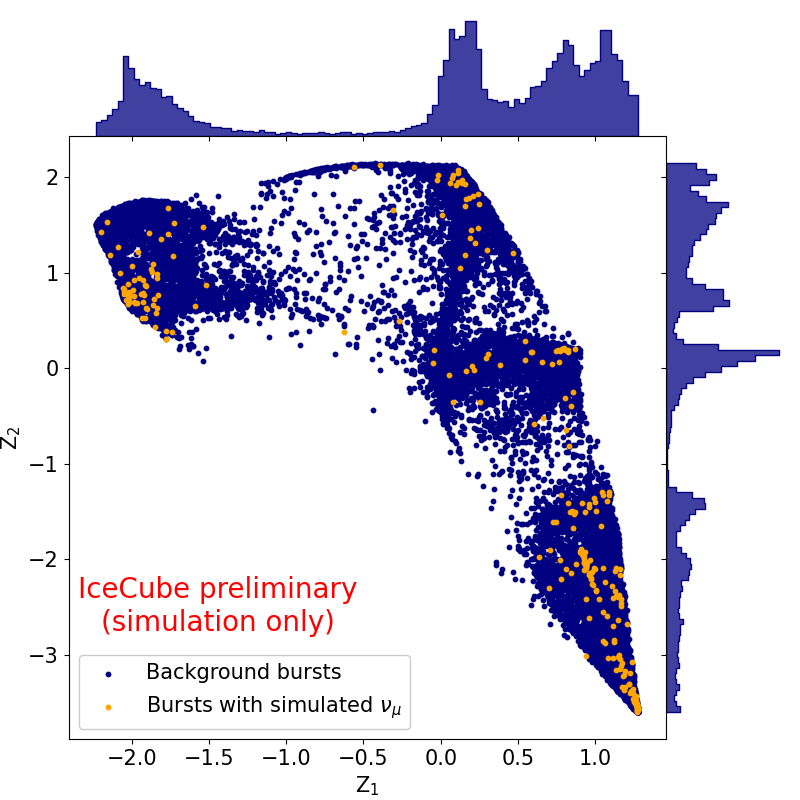}
    \includegraphics[width=0.51\linewidth]{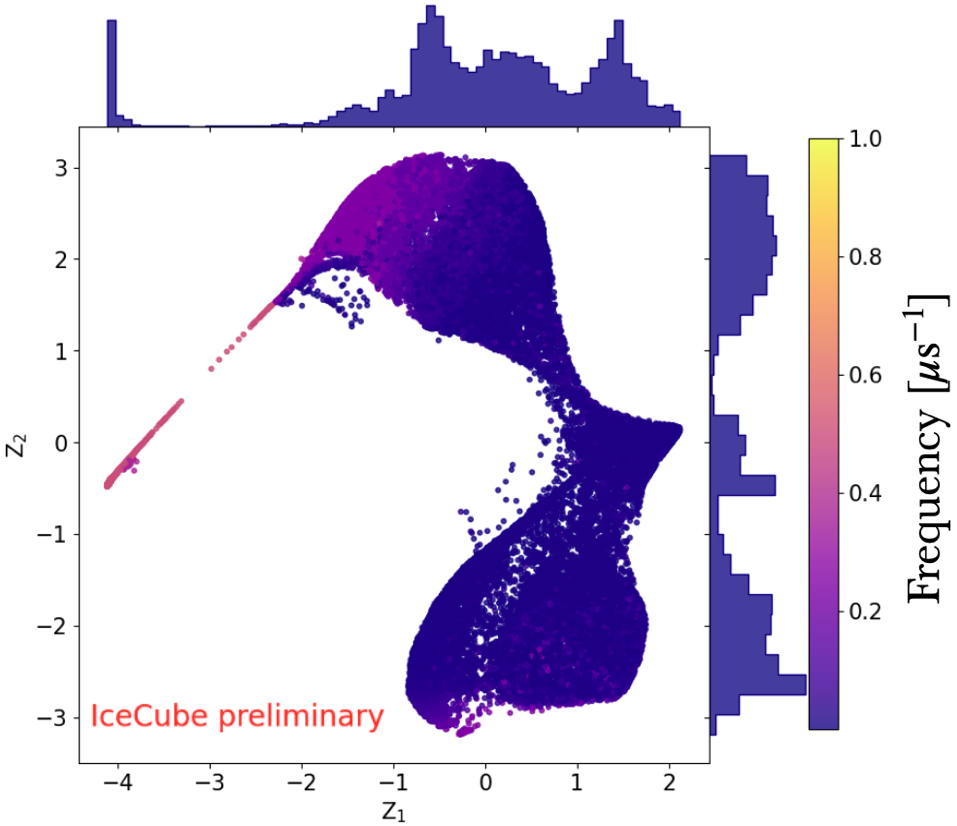}
        \caption{\textit{Left:} t-SNE distribution of bursts projected along the axis that most highlights the distinct groups of bursts. The blue points correspond to bursts constructed from background simulation only, and the orange points are bursts containing simulated muon neutrinos. \textit{Right:} t-SNE plot from data-only events. The colour shows the frequency of the burst, i.e., how many hits per microsecond. The histograms show the distribution and density of points along the individual axis.}
    \label{fig:tsne}
\end{figure}

To look for more complicated, non-linear correlations within the burst variables, we adapt a non-linear dimensionality reduction technique called t-distributed stochastic neighbour embedding (t-SNE) \cite{JMLR:v9:vandermaaten08a} to search for correlations in the different burst parameters to differentiate between neutrinos and noise. t-SNE is a manifold learning algorithm that starts by constructing Gaussian PDFs of pairwise similar points in a high-dimensional space. The goal of the algorithm is to learn a mapping from a lower-dimensional parameter space that reflects the (dis)similarity between the initial points. This is done by constructing heavy-tailed Student-t distributions in the low-dimensional parameter space to evaluate the distance between points therein. Finally, the location of the bursts in this reduced parameter space is found by minimising the Kullback-Leibler divergence between the two distributions through gradient descent. This results in a dimensionally reduced representation of our burst parameters where the original density of bursts is conserved. The resulting distributions are shown in Figure \ref{fig:tsne}.

Figure \ref{fig:tsne} demonstrates that the resulting distributions of bursts after the t-SNE algorithm show some structure. In the left-hand plot, bursts containing background only are shown in blue points, whereas bursts containing simulated neutrinos are shown in orange. Despite there being regions with a higher density of points, there seem to be no preferred region of this parameter space for the bursts including neutrinos. The distribution of orange points follows the distribution of blue points. The same t-SNE algorithm was applied to HitSpool data, and is shown in the right-hand plot of Figure \ref{fig:tsne}. Since t-SNE is a stochastic algorithm, the two distributions are not identical, but the general behaviour is similar. For the data-only case, a distribution of bursts into regions with higher density is also seen. The colour of the right plot corresponds to the frequency of the bursts, and we see that this parameter is important for the distributions of points in the upper-left part of the parameter space. Thus, the frequency of hits within a burst could be used to distinguish different sub-threshold signals after the burst-search algorithm.
 
Finally, to determine how many different clusters of points we find after dimensionally reducing our parameter space of bursts, we apply a K-Means clustering algorithm to the reduced representation \cite{8ddb7f85-9a8c-3829-b04e-0476a67eb0fd}. K-Means partitions individual bursts into $k$ sets by minimising the intra-cluster variance (i.e., the sum of squares between points close to each other). To find the optimal number of clusters describing our reduced burst parameter space, we use the Silhouette score defined as

\begin{equation}
    {\rm Silhouette\;score}=\frac{1}{N}\sum_{i=1}^N\frac{b(i)-a(i)}{\max \{ a(i),b(i) \}},
\end{equation}
where $a(i)$ is a measure of how well point $i$ belongs to its own cluster and $b(i)$ is the smallest mean distance of point $i$ to all points in other clusters. We find a maximum silhouette score when $k=3$ for both simulations and HitSpool data. This approach shows that there is structure in sub-threshold IceCube data that is otherwise discarded. From the left-hand plot of Figure \ref{fig:tsne}, we see that the preliminary implementation of t-SNE does not clearly separate the neutrino signal from the background. This is unsurprising, as the burst-search algorithm was developed on simulations that significantly differ from HitSpool data. Recently, new sub-threshold simulations have been produced, and both the burst-search algorithm and subsequent analysis are currently under revision.

\section{Conclusions and prospects}

We have presented a new pipeline allowing for processing of sub-threshold data in the IceCube Neutrino Observatory. By utilising HitSpool data, we are able to save low-level data otherwise discarded. Based on this sub-threshold data, we have developed a burst-search algorithm to identify collections of hits characteristic of $\mathcal{O}$(GeV) neutrinos. By applying dimensionality reduction and manifold learning, different subpopulations appear as distinct clusters in both simulated and real data, and additional work is ongoing to identify the physical interpretation of these clusters. With a new set of sub-threshold simulations, additional labelled machine learning approaches
will be explored to construct a new event selection in IceCube targeting sub-threshold neutrinos from transient events.

\bibliographystyle{ICRC}
\bibliography{references}

page

\section*{Full Author List: IceCube Collaboration}

\scriptsize
\noindent
R. Abbasi$^{16}$,
M. Ackermann$^{63}$,
J. Adams$^{17}$,
S. K. Agarwalla$^{39,\: {\rm a}}$,
J. A. Aguilar$^{10}$,
M. Ahlers$^{21}$,
J.M. Alameddine$^{22}$,
S. Ali$^{35}$,
N. M. Amin$^{43}$,
K. Andeen$^{41}$
% ...
\\
\\
% ...
$^{10}$ Universit{\'e} Libre de Bruxelles, Science Faculty CP230, B-1050 Brussels, Belgium \\
% ...
$^{16}$ Department of Physics, Loyola University Chicago, Chicago, IL 60660, USA \\
$^{17}$ Dept. of Physics and Astronomy, University of Canterbury, Private Bag 4800, Christchurch, New Zealand \\
% ...
$^{21}$ Niels Bohr Institute, University of Copenhagen, DK-2100 Copenhagen, Denmark \\
$^{22}$ Dept. of Physics, TU Dortmund University, D-44221 Dortmund, Germany \\
% ...
$^{35}$ Dept. of Physics and Astronomy, University of Kansas, Lawrence, KS 66045, USA \\
% ...
$^{39}$ Dept. of Physics and Wisconsin IceCube Particle Astrophysics Center, University of Wisconsin{\textemdash}Madison, Madison, WI 53706, USA \\
% ...
$^{41}$ Department of Physics, Marquette University, Milwaukee, WI 53201, USA \\
% ...
$^{43}$ Bartol Research Institute and Dept. of Physics and Astronomy, University of Delaware, Newark, DE 19716, USA \\
% ...
$^{63}$ Deutsches Elektronen-Synchrotron DESY, Platanenallee 6, D-15738 Zeuthen, Germany \\
$^{\rm a}$ also at Institute of Physics, Sachivalaya Marg, Sainik School Post, Bhubaneswar 751005, India \\
% ...

\subsection*{Acknowledgments}

\noindent
The authors gratefully acknowledge the support from the following agencies and institutions:
USA {\textendash} U.S. National Science Foundation-Office of Polar Programs,
U.S. National Science Foundation-Physics Division,
U.S. National Science Foundation-EPSCoR,
U.S. National Science Foundation-Office of Advanced Cyberinfrastructure,
Wisconsin Alumni Research Foundation,
Center for High Throughput Computing (CHTC) at the University of Wisconsin{\textendash}Madison,
Open Science Grid (OSG),
Partnership to Advance Throughput Computing (PATh),
Advanced Cyberinfrastructure Coordination Ecosystem: Services {\&} Support (ACCESS),
Frontera and Ranch computing project at the Texas Advanced Computing Center,
U.S. Department of Energy-National Energy Research Scientific Computing Center,
Particle astrophysics research computing center at the University of Maryland,
Institute for Cyber-Enabled Research at Michigan State University,
Astroparticle physics computational facility at Marquette University,
NVIDIA Corporation,
and Google Cloud Platform;
Belgium {\textendash} Funds for Scientific Research (FRS-FNRS and FWO),
FWO Odysseus and Big Science programmes,
and Belgian Federal Science Policy Office (Belspo);
Germany {\textendash} Bundesministerium f{\"u}r Forschung, Technologie und Raumfahrt (BMFTR),
Deutsche Forschungsgemeinschaft (DFG),
Helmholtz Alliance for Astroparticle Physics (HAP),
Initiative and Networking Fund of the Helmholtz Association,
Deutsches Elektronen Synchrotron (DESY),
and High Performance Computing cluster of the RWTH Aachen;
Sweden {\textendash} Swedish Research Council,
Swedish Polar Research Secretariat,
Swedish National Infrastructure for Computing (SNIC),
and Knut and Alice Wallenberg Foundation;
European Union {\textendash} EGI Advanced Computing for research;
Australia {\textendash} Australian Research Council;
Canada {\textendash} Natural Sciences and Engineering Research Council of Canada,
Calcul Qu{\'e}bec, Compute Ontario, Canada Foundation for Innovation, WestGrid, and Digital Research Alliance of Canada;
Denmark {\textendash} Villum Fonden, Carlsberg Foundation, and European Commission;
New Zealand {\textendash} Marsden Fund;
Japan {\textendash} Japan Society for Promotion of Science (JSPS)
and Institute for Global Prominent Research (IGPR) of Chiba University;
Korea {\textendash} National Research Foundation of Korea (NRF);
Switzerland {\textendash} Swiss National Science Foundation (SNSF).

\end{document}